\documentstyle[sprocl]{article}

\input{epsf}
\def\Journal#1#2#3#4{{#1} {\bf #2}, #3 (#4)}

\def\NPB{{\em Nucl.\ Phys.\ }B}
\def\PLB{{\em Phys.\ Lett.\ }B}
\def\PRL{\em Phys.\ Rev.\ Lett.\ }
\def\PRD{{\em Phys.\ Rev.\ }D}

\def\be{\begin{equation}}
\def\ee{\end{equation}}
\def\bea{\begin{eqnarray}}
\def\eea{\end{eqnarray}}

\def\nc{N_c}
\def\ln{$N_c \rightarrow \infty$}
\def\bra#1{\left\langle #1 \right | }
\def\ket#1{\left | #1 \right\rangle }
\def\Tr{{\rm Tr}}
\def\CO{{\cal O}}\def\CM{{\cal M}}\def\CP{{\cal P}}

\begin{document}
\title{QCD FROM A LARGE N PERSPECTIVE\thinspace\footnote{Plenary talk at the
XIV International Conference on Particles and Nuclei, Williamsburg,
22--28 May 1996 (PANIC 96).}}
\author{ANEESH V.~MANOHAR}
\address{Department of Physics 0319, University of California at San Diego,
\hbox{9500 Gilman Drive, La Jolla, CA 92093, USA}}


\maketitle
\abstracts{The $1/N_c$ expansion of QCD can be used to calculate properties
of nucleons and $\Delta$ baryons, such as masses, magnetic moments,
and pion couplings. The predictions of the $1/N_c$
expansion are in excellent agreement with the experimental
data. The $1/N_c$ expansion also provides an understanding of the relation
between the quark model, the Skyrme model, and QCD.
This talk reviews some of the recent developments.
}

\section{Introduction}\label{sec:intro}

Many features of the baryon sector of QCD have recently been understood
both qualitatively and quantitatively using the $1/N_c$
expansion.\cite{dm}$^{\!-\,}$\cite{djm2}
Here $N_c$
is the number of colors, which is three in the real world. The main results
are:
\begin{itemize}
\item{}
There is a new symmetry in the baryon sector of QCD in the large $N_c$
limit. For two light flavors, the symmetry is a contracted $SU(4)$ symmetry
which connects the $u\uparrow$, $u\downarrow$, $d\uparrow$, $d\downarrow$
in a baryon.
The $SU(4)_c$ symmetry relates the nucleon and $\Delta$,
which belong to a single irreducible representation of $SU(4)_c$.
\item{}
One can compute the higher order $1/N_c$ corrections in a systematic way.
\item{}
Many results previously obtained in models such as the quark model or
the Skyrme model can be proved to be true in QCD to
order $1/N_c$ or order $1/N_c^2$.
The $1/N_c$ expansion explains why some model predictions work better than
others; results obtained to order $1/N_c^2$ work about three times better than
those obtained to order $1/N_c$. The $1/N_c$ expansion provides a unified
understanding of different hadronic models, and more importantly, allows one
to compute some quantities (e.g. baryon masses, pion couplings,
magnetic moments) directly from QCD without any model assumptions.
\item{} The $1/N_c$ expansion is intimately connected with baryon chiral
perturbation theory. It explains why the $\Delta$ must be included to have
a consistent perturbative chiral expansion for baryons. Including the
$\Delta$ converts baryon chiral
perturbation theory from a strong coupling expansion in powers of $N_c$ to
a weak coupling expansion in powers of $1/N_c$.
\item{} The sigma term puzzle, and the success of
various baryon mass formul\ae\ such as the Gell-Mann--Okubo formula can now
be understood
\item{} The $1/N_c$ expansion provides some information on the nucleon-nucleon
potential.\cite{dbk} (See the talk by D.B.~Kaplan in this proceedings.)
\end{itemize}

\section{Large $N_c$ Counting Rules}\label{sec:count}

Large $N_c$ QCD~\cite{thooft} is a non-trivial theory with confinement and
chiral symmetry breaking. The physical states are mesons and baryons.
The $1/N_c$ counting rules in the meson sector are
well-known.\cite{thooft,coleman} The pion decay constant $f_\pi$,
which is like a one-meson amplitude, is of order $\sqrt N_c$.
Every additional meson in a vertex
brings a suppression factor of $1/\sqrt N_c$.
Thus, three-meson
amplitudes are of order $1/\sqrt N_c$, four-meson amplitudes are of order
$1/N_c$, etc.  The counting rules imply that
while large $N_c$ QCD is a strongly interacting
theory in terms of quarks and gluons, it is equivalent to a weakly
interacting theory of mesons (and, as we will soon see, baryons). Loop
corrections in the meson theory are suppressed by powers of $1/N_c$, which
is consistent with a semiclassical expansion in powers of $\hbar/N_c$.

Large $N_c$ baryons are more difficult to study than mesons, because the
number of quarks in the baryon is $N_c$. The large $N_c$ counting rules
for baryon amplitudes were given by Witten.\cite{witten} The baryon
matrix elements of any quark bilinear such as $\bar\psi \psi$ are
$\le \CO\left( N_c \right)$,
because the operator can be inserted on $N_c$ possible quark lines.
One finds an inequality, rather than an equality, because it is possible that
there is a cancellation between the various diagrams. These possible
cancellations will turn out to be very important in the analysis that follows.
I will assume that the baryon mass $M$ and the axial coupling $g_A$ are both
of order $N_c$, i.e. that there is no cancellation for these two quantities.
All models of the baryon, such as the Skyrme model, the non-relativistic
quark model, or the bag model, satisfy this assumption. For example,
in the non-relativistic
quark model $g_A=(N_c+2)/3$, which gives the well-known value of $5/3$ for
$N_c=3$.

With these assumptions, the pion-nucleon coupling is of the form
$$
(g_A/f_\pi)\ \bar N \gamma_\mu \gamma_5 \tau^a N \ \partial_\mu \pi^a,
$$
and is of order $\sqrt N_c$, since $g_A \sim N_c$ and $f_\pi \sim \sqrt N_c$.
Recoil effects are of order $1/N_c$, and can be neglected. This allows one
to simplify the expression for the nucleon axial current. The time component
of the axial current between two nucleons at rest vanishes.
The space components of the axial current between nucleons at rest can
be written as
\begin{equation}\label{axialmatrix}
\bra{N} \bar\psi\gamma^i\gamma_5 \tau^a\psi \ket{N} = g \nc \bra{N} X^{ia}
\ket{N},
\end{equation}
where $\bra{N} X^{ia} \ket{N}$ and $g$ are of order one.
The coupling
constant $g$ has been factored out so that the normalization of $X^{ia}$
can be chosen conveniently. $X^{ia}$ is an operator (or $4\times4$
matrix) defined on nucleon states $p\uparrow$, $p\downarrow$,
$n\uparrow$, $n\downarrow$, which has a finite $\nc \rightarrow \infty$
limit.

\section{Consistency Conditions}\label{sec:cons}

One can use the qualitative $1/N_c$ counting rules
to obtain consistency conditions for various baryonic quantities, such as the
axial couplings $X^{ia}$. These consistency
conditions can be solved to obtain results for baryons that are
in good agreement with the experimental data at $N_c=3$.

\begin{figure}
\moveright2cm\hbox{\epsfxsize=8 cm
\epsffile{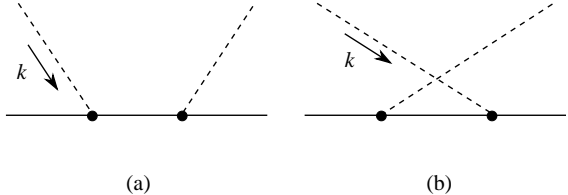}}
\caption{The leading contribution to $\pi N$ scattering to order $E$ in the
large $N_c$ and chiral limits.\label{fig:graph}}
\end{figure}

Consider pion-baryon scattering at fixed energy $E$
(as $N_c \rightarrow \infty$)
in the chiral limit. The leading contribution is from the
pole graphs in Fig.~\ref{fig:graph}, which
contribute at order $E$  provided the intermediate
state is degenerate with the initial and final states.
Otherwise, the pole graph contribution is of order $E^2$.
In the large $N_c$ limit, the pole graphs are of order $N_c$.
There is also a direct two-pion-$N$ coupling that contributes at order $E$,
which is of order $1/N_c$ in the large $N_c$ limit and can be neglected.

The
pion-nucleon scattering amplitude for $\pi^a(q) + N (k) \rightarrow
\pi^b(q^\prime)+ N(k^\prime)$ from the pole graphs is
\begin{equation}\label{amp}
-i\ q^i q^{\prime\,j} {\nc^2g^2\over f_\pi^2}
\left[{ 1\over q^0} X^{jb }X^{ia}
- {1 \over q^{\prime\,0} } X^{ia}X^{jb}\right],
\end{equation}
where the amplitude is written in the form of an operator acting on nucleon
states. Both initial and final nucleons are on-shell, so $q^0=q^{\prime\,0}$.
The product of the $X$'s in Eq.~\ref{amp}
then sums over the possible spins and
isospins of the intermediate nucleon. Since $f_\pi\sim\sqrt\nc$, the overall
amplitude is of order $\nc$, which violates unitarity at fixed energy,
and also contradicts the
large $N_c$ counting rules of Witten. Thus \ln\ QCD with a $I=J=1/2$
nucleon multiplet interacting with a pion is an inconsistent field
theory. There must be other states that cancel the order $\nc$ amplitude
in Eq.~\ref{amp} so that the total amplitude is order one, consistent
with unitarity. One can then generalize $X^{ia}$ to be an operator on
this degenerate set of baryon states, with matrix elements equal to the
corresponding axial current matrix elements. With this generalization,
the form of Eq.~\ref{amp} is unchanged, and we obtain the first consistency
condition for baryons,\cite{dm}
\begin{equation}\label{consi}
\left[X^{ia},X^{jb}\right]=0 \ .
\end{equation}
This consistency condition implies that
the baryon axial currents are represented by a set of operators
$X^{ia}$ which commute in the \ln\ limit, a result
also derived by Gervais and Sakita.~\cite{gs} There are
additional commutation relations,
\begin{eqnarray}\label{xjcomm}
\left[J^i, X^{jb}\right] &=& i\,\epsilon_{ijk}\, X^{kb}, \\
\left[I^a, X^{jb}\right] &=& i\,\epsilon_{abc}\, X^{jc}, \nonumber
\end{eqnarray}
since $X^{ia}$ has spin one and isospin one.

The algebra in Eqs.~\ref{consi} and \ref{xjcomm} is a contracted SU(4) algebra.
To see this, consider the algebra of operators in the non-relativistic quark
model, which has an $SU(4)$ symmetry. The operators are
\begin{equation}
J^i = q^\dagger {\sigma^i\over2}q,\ \
I^a = q^\dagger {\tau^a\over2}q,\ \
G^{ia} = q^\dagger {\sigma^i\tau^a\over4}q,
\end{equation}
where $J^i$ is the spin, $I^a$ is the isospin, and $G^{ia}$ are the spin-flavor
generators. The commutation relations involving $G^{ia}$ are
\begin{eqnarray}
\left[G^{ia},G^{jb}\right] &=& \frac i 4\, \epsilon_{ijk} \delta_{ab}\, J^k +
\frac i 4\,\epsilon_{abc} \delta_{ij}\, I^c,\nonumber \\
\left[J^i,G^{jb}\right] &=& i\,\epsilon_{ijk}\, G^{kb},\label{sufour} \\
\left[I^a,G^{ib}\right]  &=& i\,\epsilon_{abc}\, G^{jc}, \nonumber
\end{eqnarray}
The algebra for
\ln\ baryons in QCD is given by taking the limit
\begin{equation}\label{xlimit}
X^{ia} \equiv \lim_{\nc\rightarrow\infty} {G^{ia} \over \nc},
\end{equation}
The $SU(4)$ commutation relations Eq.~\ref{sufour}
turn into the commutation relations Eqs.~\ref{consi}--\ref{xjcomm}
in the \ln\ limit. The limiting process Eq.~\ref{xlimit} is a Lie algebra
contraction.

We have just proved that the
the \ln\ limit of QCD has a contracted $SU(4)$ symmetry in the baryon
sector. The unitary irreducible representations of the
contracted Lie algebra can be  obtained using the theory of induced
representations, and can be shown to be infinite dimensional. The simplest
irreducible representation is a tower of states with $I=J=1/2, 3/2,
etc.$, which is the set of states of the Skyrme model, or the large $N_c$
non-relativistic quark model. More complicated irreducible representations
correspond to baryons containing strange quarks.

The pion-baryon coupling matrix $X^{ia}$ can be completely determined (up to
an overall normalization $g$), since it is a generator of the $SU(4)_c$
algebra. It is easy to show that the \ln\ QCD predictions for the pion-baryon
coupling ratios are the same as those obtained in the Skyrme model
or non-relativistic
quark model\cite{am}$^{\!-\,}$\cite{cvetic} in the
$N_c\rightarrow \infty$ limit, because both these models also have a
contracted $SU(4)$ symmetry
in this limit. In the Skyrme
model, the axial current in the $N_c\rightarrow\infty$ limit is
$X^{ia}\propto \Tr A \tau^i A^{-1} \tau^a$. The $X$'s
commute (and so form part of an $SU(4)_c$ algebra),
since $A$ is a $c$-number. We have already seen how the
quark model algebra reduces to $SU(4)_c$ in the large $N_c$ limit.

\section{$1/N_c$ Corrections}\label{sec:corr}

What makes the $1/N_c$ expansion for baryons interesting is that it is
possible to compute the $1/N_c$ corrections. This allows one to compute
results for the physical case $N_c=3$, rather than for the strict
$N_c=\infty$ limit, which is only of formal interest.

\begin{figure}
\moveright2cm\hbox{\epsfxsize=8 cm
\epsffile{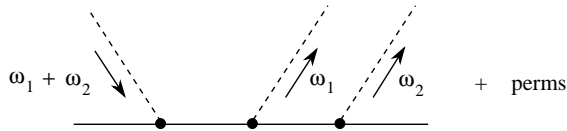}}
\caption{The leading contribution to $\pi N \rightarrow \pi \pi N$
in the large $N_c$ and chiral limits.\label{fig:three}}
\end{figure}

The $1/N_c$ corrections to the axial couplings $X^{ia}$ are determined
by considering the scattering process $\pi^a+N\rightarrow \pi^b+\pi^c+N$
at low energies. The baryon pole graphs that contribute in the \ln\
limit are shown in Fig.~\ref{fig:three}. The axial coupling $X^{ia}$ can be
expanded in a series in $1/\nc$,
\begin{equation}\label{gexp}
X^{ia}= X^{ia}_0 +{1\over\nc} X^{ia}_1+\ldots
\end{equation}
The amplitude for
pion-nucleon scattering from the diagrams in Fig.~\ref{fig:three}
is proportional to
$$
\nc^{3/2} \left[X^{ia},\left[X^{jb},X^{kc}\right]\right],
$$
and violates unitarity unless the double commutator vanishes at least as
fast as $\nc^{-3/2}$, so that the amplitude is at most of order one. (In fact,
one expects that the double commutator is of order $1/\nc^2$ since the
corrections should only involve integer powers of $1/\nc$. This result also
follows from the \ln\ counting rules which imply that each additional pion
gives
a factor of $1/\sqrt{\nc}$ in the amplitude.) Substituting Eq.~\ref{gexp}
into the constraint gives the consistency condition
\begin{equation}\label{acons}
\left[X^{ia}_0,\left[X_1^{jb},X^{kc}_0\right]\right] +
\left[X^{ia}_0,\left[X^{jb}_0,X_1^{kc}\right]\right] = 0,
\end{equation}
using $\left[X^{ia}_0,X^{jb}_0\right]=0$ from Eq.~\ref{consi}. The only
solution to Eq.~\ref{acons} is that $X_1^{ia}$ is
proportional to $X_0^{ia}$.\cite{dm} This can be verified by an explicit
computation of $X_1^{ia}$ using reduced matrix elements, or by using
group theoretic methods.\cite{djm1,djm2} Thus we find that
\begin{equation}\label{gans}
X^{ia} = \left(1 + {c\over \nc}\right) X_0^{ia} + \CO\left({1\over
\nc^2}\right),
\end{equation}
where $c$ is an unknown constant. The first correction to $X^{ia}$ is
proportional to the lowest order value $X_0^{ia}$,  so the $1/\nc$
correction to  the axial coupling constant ratios vanishes.\cite{dm}
We have now shown
that the predictions for the ratios of pion couplings, such as $g_{\pi NN}/
g_{\pi N \Delta}$ or $g_{\pi NN}/ g_{\pi \Delta \Delta}$ are equal to
the prediction in the Skyrme model up to corrections of order
$1/N_c^2$. The Skyrme model
results for the {\it ratios} of couplings
are known to agree with experiment to about 10\% accuracy.\cite{anw}

At order $1/N_c$, the baryons
in an irreducible representation of the contracted $SU(4)$ Lie algebra
are no longer degenerate, but are split by an
order $1/N_c$ mass term $\Delta M$. The intermediate baryon propagator
in Eq.~\ref{amp} should be replaced by $1/(E-\Delta M)$. The energy $E$ of
the pion is order one, whereas $\Delta M$ is of order $1/N_c$, so the
propagator can be expanded to order $1/N_c$ as
\begin{equation}\label{propexp}
{1\over E-\Delta M} = {1\over E} + {\Delta M\over E^2} +\ldots\ \ .
\end{equation}
Including the $1/N_c$ corrections to the propagator does not affect the
derivation of Eq.~\ref{consi}, as the two terms in Eq.~\ref{propexp} have
different energy dependences. The first term leads to the consistency
condition Eq.~\ref{consi} and the second gives the consistency condition on
the baryon masses,\cite{j,djm1}
\begin{equation}\label{mcons}
\left[X^{ia},\left[X^{jb},\left[X^{kc},\Delta M\right]\right]\right]=0.
\end{equation}
This constraint can be used to obtain the $1/N_c$ corrections to the
baryon masses. The constraint Eq.~\ref{mcons} is equivalent to a simpler
constraint obtained by Jenkins using chiral perturbation theory\cite{j}
\begin{equation}\label{mconsII}
\left[X^{ia},\left[X^{ia},\Delta M\right]\right]={\rm constant}.
\end{equation}
The solution of Eq.~\ref{mcons} or \ref{mconsII} is that the baryon mass
splitting $\Delta M$ must be proportional to $J^2/N_c=j(j+1)/N_c$, where
$j$ is the  spin of the baryon.\cite{j}
This is precisely the form of the baryon mass
splitting in the Skyrme model.

\begin{figure}
\setlength{\unitlength}{6mm}
\centerline{\hbox{
\begin{picture}(10,7.725)(-1.9,-0.525)
\def\level{\line(1,0){5}}
\thicklines
\put(0,6.4){\level}
\put(0,4.9){\level}
\put(0,3.6){\level}
\put(0,2.5){\level}
\put(0,1.6){\level}
\put(0,0.9){\level}
\put(0,0.4){\level}
\put(0,0.1){\level}
\put(0,0){\level}
\thinlines
\put(5.5,6.4){\line(1,0){1}}
\put(5.5,4.9){\line(1,0){1}}
\put(5.5,0.125){\line(1,0){1}}
\put(5.5,-0.025){\line(1,0){1}}
\put(-1.9,0){\line(1,0){1}}
\put(-1.9,6.4){\line(1,0){1}}
\put(6,5.65){\makebox(0,0){$1$}}
\put(7.5,0){\makebox(0,0){$1/N_c$}}
\put(-1.65,3.05){$N_c$}
\put(-1.4,3.6){\vector(0,1){2.7}}
\put(-1.4,2.8){\vector(0,-1){2.7}}
\put(6,7.2){\vector(0,-1){0.8}}
\put(6,4.1){\vector(0,1){0.8}}
\put(6,-0.525){\vector(0,1){0.5}}
\put(6,0.625){\vector(0,-1){0.5}}
\end{picture}
}}
\caption{The baryon mass spectrum including the $J^2/N_c$ term.
\label{fig:split}}
\end{figure}
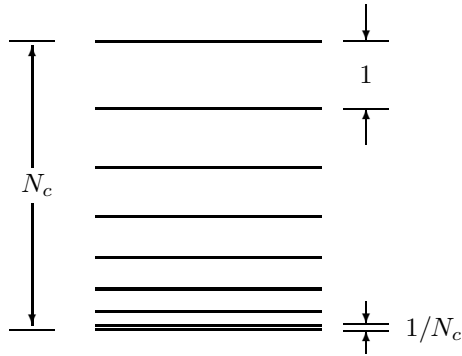

The structure of the $1/N_c$ corrections shows that the expansion parameter is
$J/N_c$, where $J$ is the spin of the baryon. For example, the baryon mass
spectrum including the $J^2/N_c$ mass splitting has the form shown in
Fig.~\ref{fig:split}.
The correction terms are only small near the bottom of the (infinite) baryon
tower. For this reason, the $1/N_c$ expansion will only be considered for
baryons with spin $J$ held fixed as $N_c \rightarrow \infty$.

\section{Results}\label{sec:res}

It is relatively simple to obtain consistency conditions such as
Eq.~\ref{consi},\ref{acons} or \ref{mcons}
using the $1/N_c$ expansion. Solving these equations is
straightforward, but will not be described here. The general solution
for any physical quantity $\CM$ has the form of a series expansion,
\begin{equation}
\CM = N_c^r\ \CP \left(X^{ia},{J^k\over N_c}, {I^a\over N_c}\right),
\end{equation}
where $\CP$ is a polynomial in its arguments, with each operator
appearing with a coefficient $c
\left( 1/N_c \right)$ which is order one as $N_c \rightarrow \infty$. The
overall power of $r$ is given by the $N_c$ power counting rule
for the quantity $\CM$ being considered. All the terms in the polynomial
$\CP$ must have the same spin and isospin as $\CM$. In addition, there are
a set of operator identities that allow one to eliminate many terms in $\CP$
with contracted spin or isospin indices.
For two flavors, these identites are~\cite{djm1}
\begin{eqnarray}
X^{ia} X^{ib} = \delta^{ab}, &\ &
X^{ia} X^{ja} = \delta^{ij},\nonumber \\
X^{ia} J^i = - I^a, &\ &
X^{ia} I^a = - J^i,\label{opiden} \\
\epsilon_{ijk} X^{ia} X^{jb} = \epsilon_{abc} X^{kc}, &\ &
\epsilon_{abc} X^{ia} X^{jb} = \epsilon_{ijk} X^{kc}.\nonumber
\end{eqnarray}
For example, if $\CM$ is the baryon mass,
$r=1$ since the baryon mass is of order $N_c$. The general expansion
has the form
\begin{equation}
M = N_c \left( c_0 + c_1 {J^2 \over N_c^2} + c_2 {J^4 \over N_c^4} +
\ldots \right).
\end{equation}
All other terms in $\CP$
can be eliminated using Eq.~\ref{opiden}. At present, it is
not known how to compute the coefficients $c_i$.

One can obtain relations among baryon quantities by working to a given order
in $1/N_c$, neglecting all higher order terms in $\CP$. The accuracy of the
relations is determined by the order in $1/N_c$ of the neglected operators.
As a trivial example, consider
the baryon masses in the two flavor case. The physical states are the $N$ and
$\Delta$, with $I=J=1/2$ and $3/2$, respectively. To lowest order, one finds
that
\begin{equation}
M_N = N_c c_0 + \CO \left(1/N_c \right),\ \ \
M_\Delta = N_c c_0 + \CO \left(1/N_c \right),
\end{equation}
and the prediction
\begin{equation}\label{rel1}
M_N =
M_\Delta + \CO \left(1/N_c \right).
\end{equation}
This relation is a homogenous relation among the baryon masses.
It is convenient to write Eq.~\ref{rel1} in the form
\begin{equation}\label{rel2}
{M_\Delta - M_N \over \left(M_\Delta + M_N \right)/2}
= \CO \left(1/N_c^2 \right),
\end{equation}
where the numerator is the difference of the two sides of Eq.~\ref{rel1},
and the denominator is the average of the two sides. The form Eq.~\ref{rel2}
does not depend on the overall scale of Eq.~\ref{rel1}.
Including the first correction gives
\begin{equation}
M_N = N_c c_0 + c_1 {3\over4 N_c}+\CO \left(1/N_c^3 \right),\
M_\Delta = N_c c_0 + c_1 {15\over4 N_c} + \CO \left(1/N_c^3 \right)
\end{equation}
In this simple example, there is no relation including the $1/N_c$ correction
because two masses are given in terms of two parameters. However, one obtains
many non-trivial predictions when the results are extended to three flavors.

\begin{figure}
\moveright1cm\hbox{\epsfxsize=10 cm
\epsffile{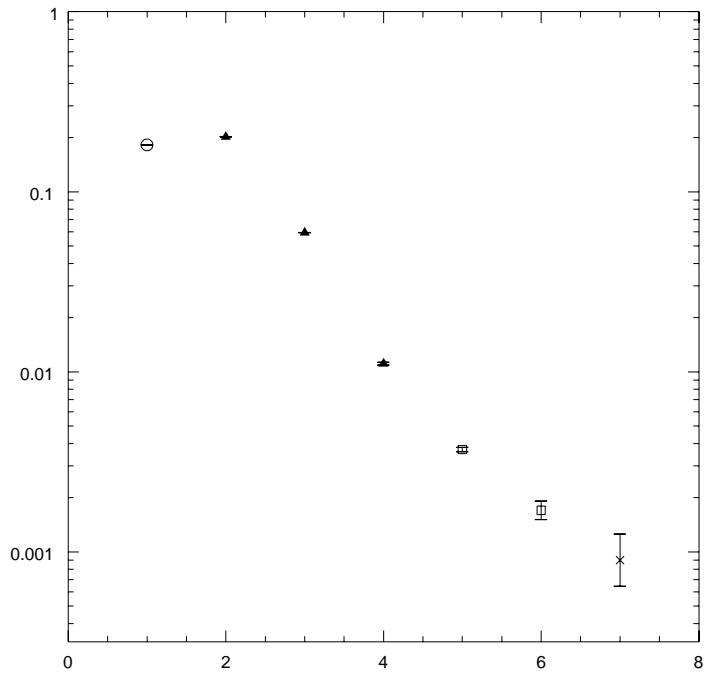}}
\caption{Plot of the isospin averaged mass relations of Jenkins and
Lebed.\protect\cite{jl}. The error bars are the experimental errors on the
measured baryon masses.
In terms of the $SU(3)$ breaking parameter $\epsilon
\sim 0.3$, and $1/N_c$,
the open circle is a relation of order $1/N_c^2$, the
three solid triangles are relations of order $\epsilon/N_c$,
$\epsilon/N_c^2$, and
$\epsilon/N_c^3$, the open squares
are relations of order $\epsilon^2/N_c^2$,
and $\epsilon^2/N_c^3$ and the $\times$ is a relation of
order $\epsilon^3/N_c^3$. The $1/N_c$ hierarchy is obvious. \label{fig:mass}}
\end{figure}

The baryon masses have been analyzed in a simultaneous expansion in $SU(3)$
breaking (due to $m_s$), isospin breaking, and $1/N_c$ by Jenkins and
Lebed.\cite{jl} They
obtain a number of relations of the form Eq.~\ref{rel2}. Their
relations for isospin averaged masses (such as $(M_P + M_N)/2$) are shown
graphically in Fig.~\ref{fig:mass}. What is plotted are mass
splittings of the form Eq.~\ref{rel2}; the error bars are the experimental
errors on the measured baryon masses.
The relations are valid to some order in $\epsilon$ and
$1/N_c$, where $\epsilon \sim 0.3$ is a measure of $SU(3)$ breaking.
It is clear from Fig.~\ref{fig:mass} that the $1/N_c$ expansion explains
the hierarchy of the baryon mass relations. For example, there are three
relations which should be true up to corrections of order $\epsilon$. One
of them is order $\epsilon/N_c$, one of order $\epsilon/N_c^2$ and one of
order $\epsilon/N_c^3$. The hierarchy between these relations is obvious from
Fig.~\ref{fig:mass}. An $SU(3)$ analysis alone does not explain why some of
the order $\epsilon$ relations work better than others.
Similarly, the $1/N_c$ expansion explains the hierarchy
in the two $\epsilon^2$ relations. One can also see that the $1/N_c^2$
mass relation, which is not suppressed by $SU(3)$ breaking,
works to about 11\%.
One can understand the baryon mass spectrum to a fractional error
of 0.001 (i.e. $\sim 1$~MeV)
in {\it the isospin averaged} masses using $1/N_c$ and $SU(3)$ symmetry.

\begin{figure}
\moveright1cm\hbox{\epsfxsize=10 cm
\epsffile{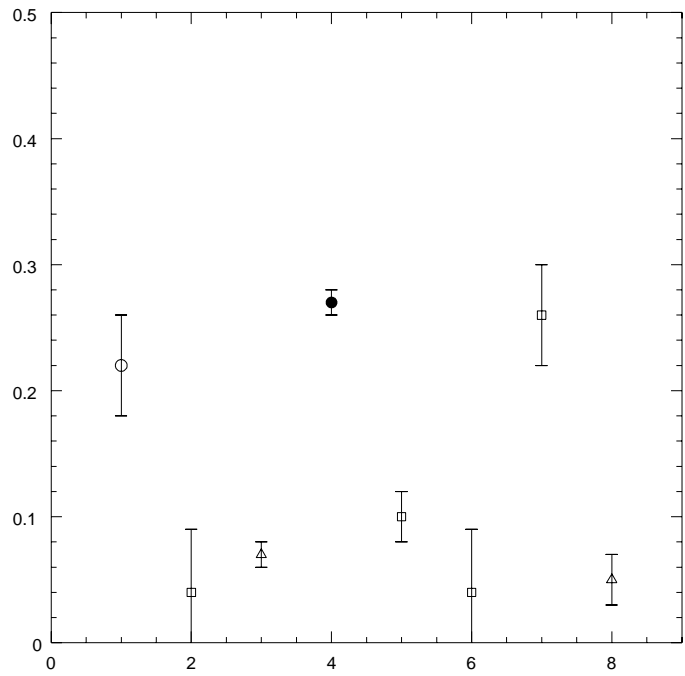}}
\caption{Plot of the magnetic moment relations. The error bars are the
experimental errors on the measured magnetic moments.
The open points are relations
which are true in $1/N_c$ and in the non-relativistic quark model, the
filled point is a relation which is true in $1/N_c$, but has no quark
model analog. The circles are relations of order $1/N_c$, the squares of
order $1/N_c^2$, and the triangles of order $\epsilon/N_c$.
\label{fig:mag}}
\end{figure}

A similar analysis can be done for the baryon magnetic
moments.\cite{jm,lmrw} The $1/N_c$ relations for the isovector and isoscalar
magnetic moments are shown in Fig.~\ref{fig:mag}. Most of the magnetic
moment relations derived using the quark model can now be obtained directly
from QCD, to some order in $1/N_c$. The $1/N_c$ expansion also explains why
some relations work better than others; something which is left unexplained
by the quark model. There is also one new relation. Relation~7, which relates
the $\Delta^+ \rightarrow p$ transition moment to $p-n$, is a $1/N_c^2$
relation
(which is also true in the quark model) which works to 30\%. This is one
instance in which the $1/N_c^2$ correction is substantial. All the $1/N_c$
relations
are homogeneous, and do not set the scale of the magnetic moments. In the quark
model, one can get a good estimate of the size of the magnetic moments from
the constituent quark mass. This is not explained by the $1/N_c$ expansion.

\section{Other Results}

There are many other results that have been obtained using the $1/N_c$
expansion, some of which are summarized here.
\begin{itemize}
\item{} One loop corrections to the baryon axial currents are large,
and disagree with experiment,
unless one includes intermediate $\Delta$ states in loops.\cite{jmchpt}
There are large cancellations between $N$ and $\Delta$ intermediate states.
The best fit values for the axial couplings when $\Delta$'s are
included are close to $SU(6)$
values.\cite{jmchpt} The $1/N_c$ explains all of these results. One
can show that the one-loop
corrections to the axial currents have the structure
\begin{equation}\label{chcancel}
N_c \left[X,\left[X,X\right]\right]
\end{equation}
Eq.~\ref{chcancel} is a commutator that is constrained by
the large $N_c$ consistency conditions, and is of order $1/N_c^2$. Thus
including intermediate $\Delta$ states produces a cancellation to two orders
in $1/N_c$, so the one-loop correction is of order $1/N_c$. This cancellation
converts an expansion in $N_c$ to a semiclassical
expansion in $\hbar/N_c$.
If intermediate $\Delta$ states are not included, the one-loop correction is
of order $N_c$ times the tree level result, since the cancellation conditions
on $X$ do not hold in the nucleon sector alone. Baryon chiral perturbation
theory without a $\Delta$ is a strong coupling expansion in powers of $N_c$.
The strongly coupled theory presumably dynamically generates a $\Delta$.
\item{}
One loop corrections to the baryon masses are large
($\sim$~500 MeV). Most of the correction is $SU(3)$ singlet or octet,
so the large non-linearities do not violate the Gell-Mann--Okubo
formula.\cite{j1992} This resolves the sigma-term
puzzle.\cite{j1980}
\item{} Certain Skyrme model results on baryons containing heavy
quarks using the bound state approach to strangeness
of Callan and Klebanov follow
from large $\nc$ alone.\cite{j,ck,jmw}
\item{} The chiral bag model arises naturally in the
$\nc\rightarrow \infty $ limit.\cite{dhm,am94}
\item{} One can compute the $F/D$ ratio and an equal spacing rule for the axial
couplings in $\beta$ decay.\cite{djm1}
\item{} One understands the
$s$-wave hyperon non-leptonic decay amplitudes.\cite{ejhnld,djm2}
\item{} One can analyze $SU(3)$ breaking in the baryon axial vector
currents.\cite{dai} One finds that $\Delta u + \Delta d - 2 \Delta s$
is shifted
from its $SU(3)$ symmetric value of $0.6$ to $0.3 \pm 0.1$. This affects the
interpretation of the recent experiments on the $g_1$ structure function of
the nucleon.

\end{itemize}

\section{Conclusions}

The $1/N_c$ expansion shows  that most of the spin-flavor structure of
baryons can be understood using contracted $SU(4)$ symmetry. It
provides a unifying symmetry which connects the
various quark models (non-relativistic, bag), the Skyrme model, and QCD.
The $1/N_c$ expansion
explains many results found earlier in baryon chiral perturbation
theory in a systematic way, and it is possible to combine
baryon chiral perturbation theory with the $1/N_c$ expansion of
QCD.\cite{ejln}

\section*{Acknowledgments}

This work was supported in part by the Department of Energy
grant DOE-FG03-90ER-49546, by a
PYI award PHY-8958081 from the National Science Foundation,
and by a ``Profesor Visitante IBERDROLA de Ciencia y Tecnolog\'{\i}a''
position at
the Departmento de F\'{\i}sica Te\'orica of the University of Val\`encia.
I would like to thank the Benasque Center for Physics for hospitality while
this manuscript was written.

\end{document}